\documentclass[12pt, a4paper, oneside]{article}
\usepackage[T1]{fontenc}
\usepackage[utf8]{inputenc}
\usepackage{graphicx}
\usepackage{bbold}
\usepackage{amssymb}
\usepackage{algpseudocode,algorithm,algorithmicx}
\usepackage[affil-it]{authblk}
\usepackage{subfig}

\usepackage{amsthm}
\theoremstyle{definition}
\newtheorem{definition}{Definition}[section]
\theoremstyle{example}

\theoremstyle{remark}

\title{The use of the multi-cumulant tensor analysis for the algorithmic 
optimisation of investment portfolios.}

\author{Krzysztof Domino$^{^\dagger}$ \texttt{kdomino@iitis.pl}}
\affil{$^{^\dagger}$ Institute of Theoretical and Applied Informatics, Polish 
Academy of Sciences, Gliwice, Poland}

\begin{document}
\maketitle

\section*{Abstract}
The cumulant analysis plays an important role in
non Gaussian distributed data analysis. The shares' prices returns are good 
example of such data. The purpose of this research is to 
develop the cumulant based algorithm and use it to determine eigenvectors that 
represent investment portfolios with low variability. Such algorithm is based 
on the Alternating Least Square method and involves the 
simultaneous minimisation $2$'nd -- $6$'th 
cumulants of the multidimensional random variable (percentage shares' returns 
of many companies). Then the algorithm was tested during the recent crash on 
the Warsaw Stock Exchange. To determine incoming crash and provide enter and 
exit signal for the investment strategy the Hurst exponent was calculated using 
the local DFA. It was shown that introduced algorithm is on average better that 
benchmark and other portfolio determination methods, but only within 
examination window determined by low values of the Hurst exponent. Remark 
that the algorithm of is based on cumulant tensors up to the $6$'th order 
calculated for a multidimensional random variable, what 
is the novel idea. It can be expected that the algorithm would be useful in the 
financial data analysis on the world wide scale as well as in the analysis of 
other types of non Gaussian distributed data. 

\paragraph{Keywords} cumulant tensors, ALS--class algorithm, Hurst exponent, 
financial data analysis, stock exchange. 

\section{Introduction}
Let us consider the  multidimensional frequency distribution of shares' prices'
percentage returns. The optimization (minimization) of higher cumulants of this
distribution is used to determine investment portfolios, to test if they are 
better on  average than the benchmark, during the crash. 
The proposed procedure is based on \cite{morton2009algebraic} and
implies the investigation of cumulants tensors -- the $n$'th cumulant of 
the 
multidimensional 
random variable is represented by the $n$--dimensional tensor 
\cite{morton2009algebraic, 
kolda2009tensor}. For this purpose, I introduce the generalisation of the 
classical 
Value at Risk (VaR) procedure \cite{best2000implementing}, where the  
left Eigenvector Decomposition (EVD) of the second cumulant 
(the covariance) matrix is performed, and the multidimensional normal 
distribution of financial data is assumed. In classical EVD approach, the 
portfolio with minimal variance corresponds to the last 
eigenvector. However, the classical EVD method fails to anticipate the risk of 
investment 
portfolios 
since the
second cumulant fails to represent the extreme events, where drops of shares'
prices values are high and cross--correlated. This happens mainly due to the 
break down of the central limit theorem resulting from the time--varying 
variance of financial data. The Autoregressive Conditional Heteroskedasticity 
(ARCH), that violates both independence and identical distribution assumptions 
of the central limit theorem,  was recorded for many types of financial data 
\cite{akgiray1989conditional, bollerslev1987conditionally, 
bollerslev1992arch, engle1995multivariate, schwert1990heteroskedasticity}. 
Recall also the impact of long 
range auto--correlations of 
shares' returns
\cite{mandelbrot1997variation, grech2008local, czarnecki2008comparison,
vasconcelos2004guided, domino2011use, domino2012use}. It is worth to mention 
the work
\cite{malevergne2002multi, rubinstein2006multi}, where authors shows that 
moments or cumulants (of order $6$ or $8$) 
may be necessary to account for the severe price fluctuations, that are usually 
observed at short time scales (e.g. portfolio rebalanced at a weekly, time 
scales). In my research I would examine portfolio rebalanced at the $20$ 
trading days (approximately monthly scale) as it is often performed in practice 
in assets management. To search for the severe price fluctuations I used the 
Hurst exponent indicator.

Following this arguments, high cumulants analysis should anticipate
extreme events, improving the search for portfolios with low variability. There 
are some works 
implying the use of $2$'nd, $3$'rd and $4$'th cumulant of 
multivariate shares' returns \cite{arismendi2014monte, jondeau2015moment}. 
In this research I use the $5$'th and the $6$'th cumulant as well, what 
is a new approach for multivariate shares' returns. In
general the proposed algorithm is based on the High Order Singular
Value Decomposition (HOSVD) and Alternating Least Square (ALS) procedure
\cite{kolda2009tensor}. 
To compare the proposed method with others (such as EVD), the author, for each 
method, creates the family of investment portfolios which are supposed to be 
safer than a benchmark. Then 
portfolios are compared using the result function that is an average percentage 
change of portfolios' values -- an average portfolio results.
Other result functions are also discussed: 
\begin{enumerate}
	\item a mode of percentage change of portfolios' values, 
	\item a maximal loss / minimal gain -- the result of the ``worst 
	portfolio'',
	\item a minimal loss/ maximal gain -- the result of the ``best portfolio''.
\end{enumerate}

The major motivation for this research is to introduce the automation method of 
analysis of data that are not Gaussian distributed. Good example of such data 
are financial data, especially during the rupture and crash period.
It is why, I focus in this work, on the financial data analysis. To determine 
the rupture and crash period and introduce the enter and exit signal of an 
investment strategy, I use the Hurst exponent indicator calculated for the 
WIG20 index, using the local DFA. This paper give some additional incentive for 
the development of cumulants tensors calculation method at low computational 
complexity. Afterwards the 
multi--cumulant analysis may be applied for large financial data sets and 
tested against many crashes on many markets. Additionally the method may be 
used to analyse other (non--financial) data that are not Gaussian distributed.

\section{The classical approach, the covariance matrix EVD}

Let us take the 
$M$--dimensional random
variable of size $T$, $\textbf{X} \in \mathbb{R}^{( T \times M)}$, being  
the percentage 
returns of $M$ shares. Its marginal variables are $X_i$, and values are 
$x_{t,i}$:
\begin{equation}\label{eq::rw}
\textbf{X} = [X_1, \ldots, X_i, \ldots, X_M] = \left[ \begin{array}{ccc}
x_{t=1, 1} & \cdots & x_{t=1, M} \\ 
\vdots & \vdots  & \vdots  \\ 
x_{t=T, 1} & \cdots & x_{t=T, M} \\ 
\end{array}   \right].
\end{equation}
An unbiased estimator of variance of the $i$'th marginal random variable 
($X_i$) is:
\begin{equation} 
\sigma^2_i = \frac{1}{T-1} \sum_{t=1}^{T}(x_{t,i} -
\overline{X_{i}})^2, 
\end{equation}
and an unbiased estimator of covariance between ($X_i$) and ($X_j$) is:
\begin{equation} 
\textrm{cov}_{i,j} = \frac{1}{T-1} \sum_{t=1}^{T}(x_{t,i} -
\overline{X_{i}})(x_{t,j} - \overline{X_{j}}).
\end{equation}\label{f::cov2}
The variance and the covariance can be represented by the $M \times M$ 
symmetric 
covariance matrix, called also the 
second cumulant matrix -- $C_2$ (notice $\sigma_i^2 = \textrm{cov}_{i,i}$):
\begin{equation}\label{f::cov3}
C_2 = \left[ \begin{array}{cccc}
\sigma^{2}_{1} & \textrm{cov}_{1,2} & \cdots & \textrm{cov}_{1,L} \\ 
\textrm{cov}_{2,1} & \sigma^{2}_{2} & \cdots & \textrm{cov}_{2,L}  \\ 
\vdots & \vdots & \ddots & \vdots  \\ 
\textrm{cov}_{L,1} & \textrm{cov}_{L,2} &\cdots & \sigma^{2}_{L} \\ 
\end{array}   \right].
\end{equation}
\begin{definition}{The Eigenvalue Decomposition -- EVD.}\label{d::EVD}
Consider the covariance (second cumulant) symmetric matrix. The matrix can be 
diagonalized in the following way:
\begin{equation}
C_2 = V \Sigma V^{\intercal}, 
\end{equation}\label{f::cov4}
where $\Sigma = V^{\intercal} C_2 V$ is the diagonal matrix with diagonal 
values  $\sigma'^2_{i} = (V^{\intercal} C_2 V)_{ii}$ and $V$ is unitary $M 
\times M$ factors matrix, such that $\sigma'^2_{i}$ are sorted in descending 
order:
\begin{equation} 
\Sigma = \left[ \begin{array}{cccc}
\sigma'^2_{1} & 0 & \cdots & 0 \\ 
0 & \sigma'^2_{2} & \cdots & 0  \\ 
\vdots & \vdots & \ddots & \vdots  \\ 
0 & 0 &\cdots & \sigma'^2_{M} \\ 
\end{array}   \right].
\end{equation}
The $i$'th column of $V$ is the eigenvector that corresponds with the 
eigenvalue 
$\sigma'^2_{i}$. Rows in the $i$'th column of $V$ are factors that give the 
linear combination of marginal random variables with the combination's variance 
$\sigma'^2_{i}$. 
The last eigenvector 
would
give the linear combination of marginal random variables with the smallest 
combination's
variance -- $\sigma'^2_{M}$.
\end{definition}

The classical EVD procedure has been often used in the portfolio risk 
determination.
However, it requires the multidimensional Gaussian distribution of 
shares' returns, where all information about the variability of the frequency 
distribution is stored in the covariance matrix. As mentioned before the 
financial data 
(shares' returns) are not Gaussian 
distributed and the classical EVD procedure has often 
failed in the investment portfolio's risk 
determination \cite{cherubini2004copula}. It is why the author proposes to 
extend the 
classical EVD procedure by taking into consideration also cumulants of order 
higher 
than $2$ -- the higher cumulants.
\section{Cumulants}
Let us consider the $M$ dimensional random variable
$\textbf{X} = [X_1, \ldots, X_M]$. The $n$'th cumulant $C_n$ of 
such variable is 
the $n$--mode tensor \cite{kolda2009tensor}, with elements 
$\kappa_{\alpha_1,\ldots,\alpha_n}(\textbf{X})$ \cite{kendall1946advanced, 
lukacs1970characteristics}:
\begin{equation}\label{eq:generating-func}
\kappa_{\alpha_1, \ldots, \alpha_n}(\textbf{X})  =
\frac{\partial^n}{\partial \tau_{\alpha_1},\partial \tau_{\alpha_2},\ldots, 
	\partial \tau_{\alpha_n}} \log\left(E\left(\exp( 
\tau\cdot\textbf{X}^{\intercal})\right) 
\right) \bigg|_{\tau = 0}.
\end{equation} 
where $\tau$ is the argument vector $\tau = \underbrace{[\tau_1, \cdots, 
\tau_i, \cdots \tau_M]}_M$, and $E()$ is the expected value operator. 
Formulas used to calculate cumulants up to $4$'th order are well known 
\cite{kendall1946advanced, lukacs1970characteristics}. The author 
has calculated $5$'th and $6$'th cumulants by the direct use of 
	(\ref{eq:generating-func}). Here analysed data were substituted for the 
	random variable 
	$\textbf{X} \in \mathbb{R}^{( T \times M)}$, and 
	computer
	differentiations were performed 
	at point $\tau = 
	\underbrace{[\tau_1, \cdots 
		\tau_M]}_M = 0$, using ForwardDiff and DualNumbers library in Julia 
		programming \cite{juliafd}.

\subsection{The multi--cumulant decomposition.}
To investigate the financial data the author takes many cumulant tensors $ 
C_2, \dots, C_n$, where $n = 4$ or $n = 6$. The calculation of cumulants 
of order $n > 6$  might require larger data series, but non--stationary of 
financial data 
\cite{grech2008local} makes the investigation of long time series less adequate 
than shorter data series. To achieve the factor matrix $V$, the author proposes 
the following ALS--class 
algorithm, where the
search for the local maximum of the function $\Phi(V)$ 
is performed \cite{de2004dimensionality, savas2010quasi}.
Following the maximisation procedure which can not be solved precisely, the 
author will
find the local maximum using the iteration procedure \cite{savas2010quasi} and
show that the results are meaningful.
\begin{definition}{The $\Phi(V)$ function.}
Consider the $i$'th core--tensor $T_i$ that is the contraction of 
$C_i$ tensor and $i$ factor matrices $V$:
\begin{equation}
(T_{i})_{l_1, \cdots, l_i} = \sum_{j_1, \cdots, j_i} (C_{i})_{j_1, \cdots, j_i} 
V_{j_1 l_1} \cdots V_{j_i l_i}.
\end{equation}
 The ALS procedure 
 proposed in 
 \cite{morton2009algebraic, 
 	savas2010quasi} refers
 to the search for the common factor matrix $V$ that maximise $\Phi_4(V)$.
\begin{equation}\label{eq::f4}
\Phi_{4}(V) = \frac{1}{2!} ||V^{\intercal} C_2 V||^2 + \sum_{i = 
3}^4\frac{1}{i!} 
||T_i||^2.
\end{equation}
The author proposes to extend the analysis up to the $6$'th cumulant which are 
more 
sensitive to extreme ``tail events''. Hence the author defines $\Phi_6(V)$:
\begin{equation}\label{eq::f5}
\Phi_6(V) = \frac{1}{2!}||V^{\intercal} C_2 V||^2 + \sum_{i = 3}^6\frac{1}{i!} 
||T_i||^2.
\end{equation}
\end{definition}
To find the common factor matrix $V$, the ALS--based algorithm is proposed by 
author and presented at subsection~(3.2). The idea of the algorithm is 
based on the algorithm proposed in 
\cite{de2004dimensionality} where the iteration procedure was used for the 
search for the local maximum of the following function:
\begin{equation}\label{eq::f3}
\Phi'(V) = ||V^{\intercal} C_2 V||^2 +  \alpha_n||T_n 
||^2.
\end{equation}
 The proposed algorithm works for the general case (any $\Phi_n(V)$), but 
computations were
performed for $n = 4$ and $n = 6$. Racall that ALS
algorithms move information into the upper left corner of the 
core--tensor and order the information in the 
sense of the Frobenius Norm. Take the linear
transformation of analysed data $\textbf{X}$:
\begin{equation}\label{eq::comb}
\textbf{Y} = \textbf{X} V,
\end{equation}
where $\textbf{Y} = [Y_1, \ldots, Y_j, \dots,  Y_M]$. Here $Y_j$ represents 
percentage returns of the $j$'th portfolio. Elements of $\textbf{Y}$ are:
\begin{equation}\label{eq::combs}
y_{t,j} = \sum_{i= 1}^M x_{t, i}V_{i,j}
\end{equation}
The rear columns of the factor matrix would 
give the investment portfolio with little variability.

\subsection{The algorithm.}
The algorithm used to determine the factor matrix $V$ given cumulant symmetric 
tensors $C_2 \cdots C_n$, it is a general algorithm and work for each $n \geq 
3$. Let ${C_i}_{(1)}$ be the unfold of the tensor $C_i$ in the first mode 
\cite{kolda2009tensor}. The first factor matrix anzatz is computed as a matrix 
that columns are left eigenvectors of the following matrix:
\begin{equation}
\left[\frac{C_2}{2!}  \cdots \frac{{C_i}_{(1)}}{i!} \cdots
	\frac{{C_n}_{(1)}}{n!}\right]
\end{equation}
At $k$'th interaction, we have the $(V_{k-1})$ factor matrix. Now the following 
procedure is performed. The contraction of the
$C_i$ tensor (matrix) and $i-1$ factor matrices $(V_{k-1})$ is performed:
\begin{equation}
(S_{i})_{j_1, l_2 \cdots, l_i} = \sum_{j_2, \cdots, j_i} (C_{i})_{j_1, j_2 
\cdots, j_i} (V_{k-1})_{j_2 l_2} \cdots(V_{k-1})_{j_i l_i}.
\end{equation}
To compute $V_k$ we takes left eigenvectors of the following matrix:
\begin{equation}
\left[\frac{S_2}{2!}  \cdots  \frac{{S_i}_{(1)}}{i!} \cdots
\frac{{S_n}_{(1)}}{n!}\right]
\end{equation}
The procedure is repeated to satisfaction the stop condition.

\section{The investigation of financial data.}
The cumulant analysis was performed in the optimal portfolios 
searching 
problem. Let us consider the price of a $i$'th share at time $t$ -- $p_{t,i}$. 
Its percentage return is
\begin{equation}
x_{t,i} = \frac{p_{t,i}-p_{(t-1),i}}{p_{(t-1),i}} \cdot 100 \%.
\end{equation}
In our case $t$ numerates trading days (the analyse of daily 
returns was performed) and $p_{t,i}$ the closing price of $i$'th share the 
given trading day numbered by
$t$. Next the multidimensional random variable $\textbf{X}$ of percentage 
returns is 
constructed.
To construct investment portfolios we use the factor matrix 
$V$. The $j$'th portfolio returns are one dimensional random variable 
$Y_j$ with elements $y_{t,j}$. 

The naive method of factor matrix determination
uses the Eigenvalue Decomposition (EVD) of the
covariance matrix \cite{best2000implementing}. This procedure is not fully 
adequate since shares returns are not
Gaussian distributed, especially the rupture and crisis period \cite{ 
mandelbrot1997variation, grech2008local, czarnecki2008comparison, 
vasconcelos2004guided} -- importantly such period can be predicted by the use 
of the Hurst 
exponent. 
To anticipate higher cumulants of shares returns as well, the author proposes to
determine the factor matrix $V$ by searching for the local maximum of the
$\Phi_4(V)$ function as well as $\Phi_6(V)$ function -- using cumulant tensors
up to the $6$'th order, what is a new approach. The proposed $\Phi_6(V)$ method 
is used to chose portfolios with returns that have low absolute values of high
cumulants. Hence the method is supposed to work well where the portfolio's 
variability is a disadvantage. It happens during the crash of the financial 
market, hence the author tests the method 
during the last rupture and crisis on the Warsaw Stock Exchange. 

\subsection{The data analysis.}

\begin{figure}
	\includegraphics{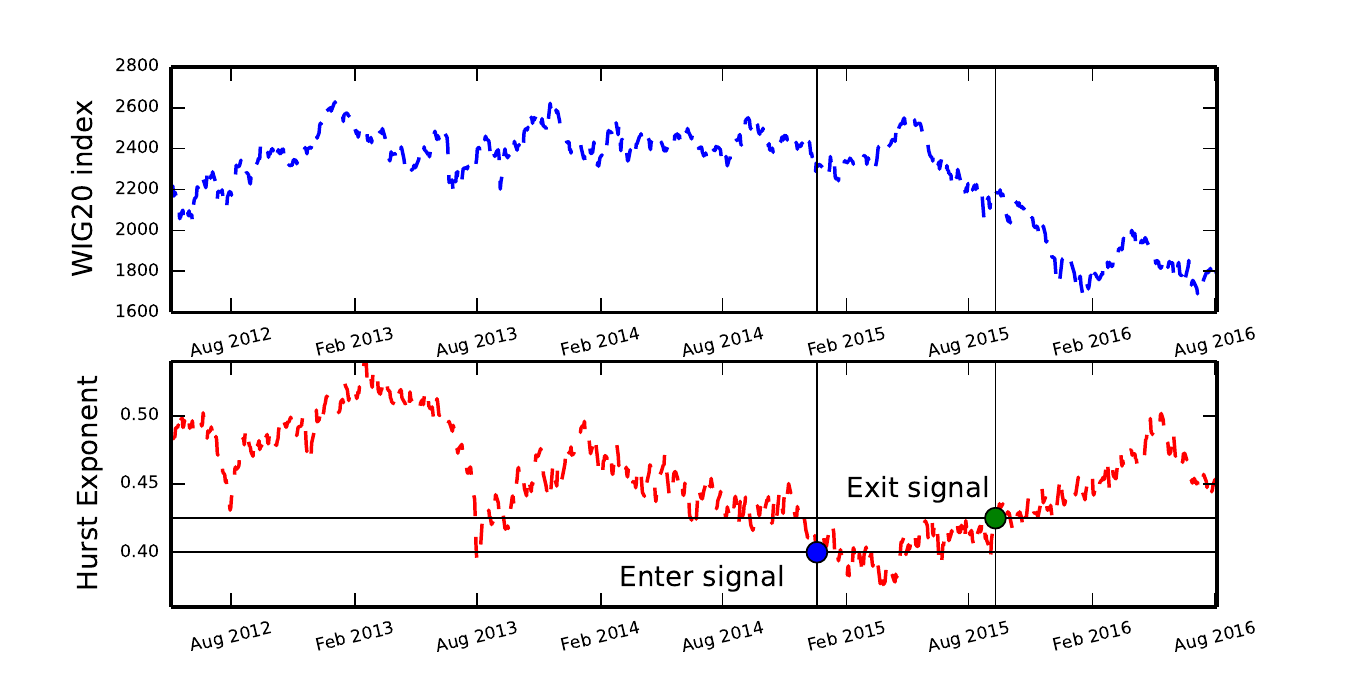}
	\caption{WIG20 index and the Hurst exponent.}
	\label{f::wh}
\end{figure}

The author has examined $M = 10$ dimensional random variable Tab.~(1),
being daily percentage returns of the shares of $10$ most liquid companies from 
the WIG20 index at the time 12.05.2010 -- 04.08.2016 (the WIG20 index includes 
$20$ most liquid companies traded on the Warsaw Stock 
Exchange). Recent composition of the WIG20 index is presented in 
Fig.~(\ref{f::wh})
The WIG20 index reached maximum at 14.05.2015 and then has fallen rapidly -- 
the crash has occurred. To introduce the signal of incoming crash, the Hurst 
exponent was calculated for the WIG20 index using the local Detrended 
Fluctuation Analysis (DFA) \cite{grech2008local, domino2011use}.
\begin{table}\label{tab:portions}
\centering
\begin{tabular}{|c|c|c|c|}
    \hline $i$ & company & contribution  & contribution  to \\ & & to WIG20   
    \% & 
     benchmark  
    \% $(BP_i)$ \\ 
    \hline 1& PKOBP & 14.64 & 18.31 \\ 
    \hline 2& PZU & 14.04 & 17.55 \\ 
    \hline 3& PEKAO & 11.65 & 14.57 \\ 
    \hline 4& PKNORLEN & 8.45 & 10.57 \\ 
    \hline 5& PGE & 7.52 & 9.40\\ 
    \hline 6& KGHM & 7.14 & 8.93 \\ 
    \hline 7& BZWBK & 5.21 & 6.51 \\ 
    \hline 8& LPP & 4.77 & 5.96 \\ 
    \hline 9& PGNIG & 3.55 & 4.43 \\ 
    \hline 10& MBANK & 3.00 & 3.75 \\ 
    \hline 
\end{tabular}
\caption{the 10 most liquid companies of the WIG20 index, their value 
contribution to the 
WIG20 index (at 20.03.2015) and as their value contribution to proposed 
benchmark 
portfolio.}
\end{table}

\subsubsection{The Hurst exponent.} To determine the rupture and crisis period 
of the stock exchange, where the examined investment strategy was tested the 
Hurst exponent was calculated using the local DFA. The parameters for DFA were 
the same as in \cite{domino2014use}: $500$ days long observation window was 
used to examine past closing value of the WIG20 index. Having the Hurst 
exponent, I introduce the 
signals of entry and exit for proposed 
investment strategy. Recall that in \cite{grech2008local} the Hurst exponent 
was calculated 
using the local DFA for the index of Polish Stock Exchange, and it was shown 
that 
before a crash (near a rupture point), the Hurst Exponent has minima $\lesssim 
0.4$. Hence the entry threshold value was chosen as $H_{entry} = 0.4$. The exit 
threshold was chosen as $H_{exit} = 0.425$ -- data with high negative 
auto--correlation was chosen for a test. 

Regarding the recent crash the entry signal 
occurred at 19.12.2014 and the exit signal at 10.09.2015. To examine the 
algorithm, I introduce the $20$ trading days (approx. 1 months) long investment
windows -- as it is performed in practice assets management. First window 
starts a day after the 
enter signal -- 22.12.2014, and there are $9$ windows within a test period 
22.12.2014 -- 27.01.2015, 27.01.2015 -- 
24.02.2015, 24.02.2015 -- 24.03.2015, 24.03.2015 -- 23.04.2015, 23.04.2015 -- 
22.05.2015, 22.05.2015 -- 22.06.2015, 22.06.2015 -- 20.07.2015, 20.07.2015 -- 
17.08.2015, 17.08.2015 -- 14.09.2015 (the last window
ends just after exit point). For each window, cumulants are 
calculated using a test series of length $T = 1100$, that ends 
just before the examination window. Next 
investment returns are analysed for data in given window -- the testing 
set. In next subsections the 
analysis is discussed in details for the $7$'th window of 22.06.2015 -- 
20.07.2015.
Then the analysis results are presented for other windows.

\subsubsection{Optimal portfolios determination -- training.}

Let us discuss in details the procedure for the exemplary window of 22.06.2015 
-- 20.07.2015.
Given the training set, the factor matrix is determined using different methods,
such as EVD, $\Phi_4(V)$ and $\Phi_6(V)$. Here also the Independent Component 
Analysis 
(ICA) was used for more general comparison. The $\Phi_4(V)$ method requires 
the 
calculation of $3$'rd and $4$'th cumulants. For $\Phi_6(V)$ also $5$'th and 
$6$'th cumulant tensors are required, which were calculated by the direct use of
Eq.~(\ref{eq:generating-func}).
Given $\Phi_4(V)$ and $\Phi_6(V)$ the algorithm introduced in subsection~(3.2) 
was 
used for the 
factor matrix $V$ 
determination.

\begin{figure}
		\subfloat[second cumulant]{
			\includegraphics{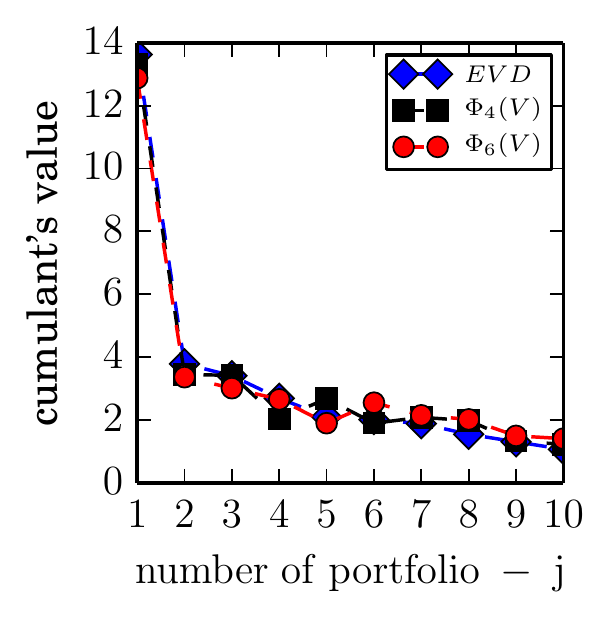}}
		\subfloat[third cumulant (normalized)]{
			\includegraphics{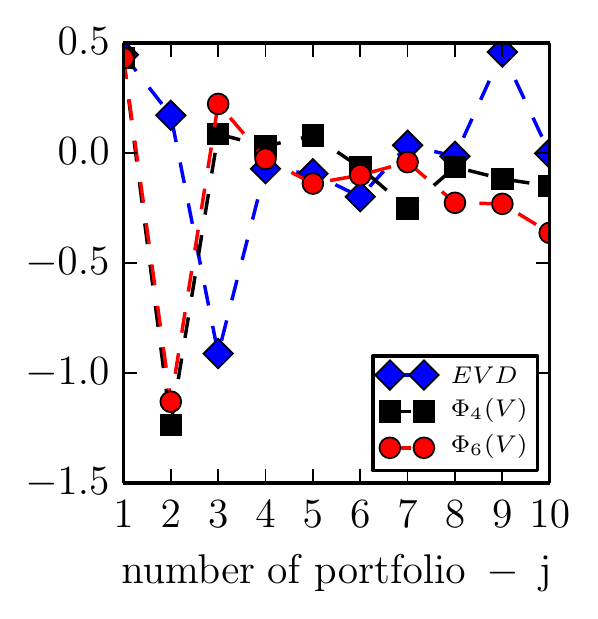}} \\	
				\subfloat[fourth cumulant (normalized)]{
					\includegraphics{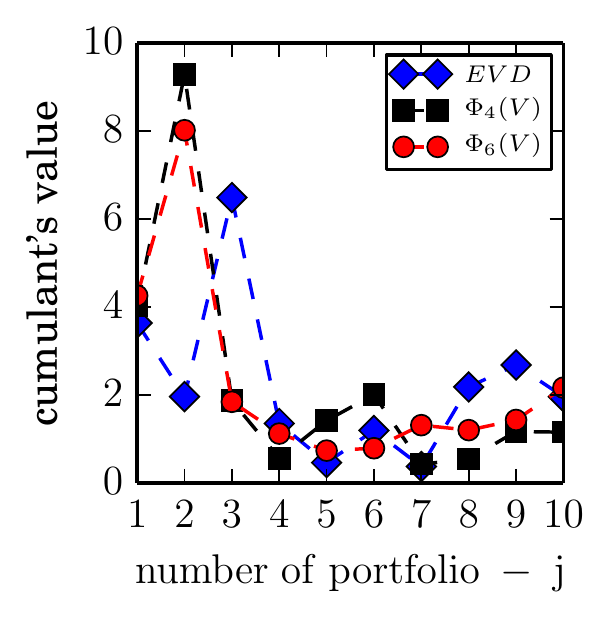}}
				\subfloat[sixth cumulant (normalized)]{
					\includegraphics{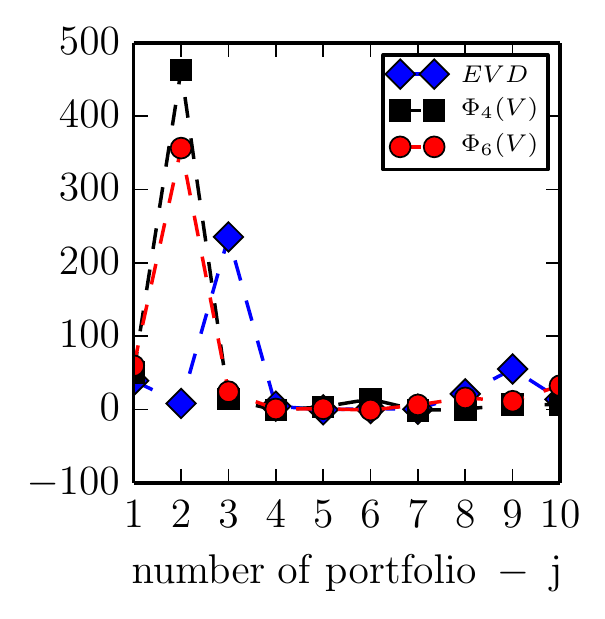}} 
	\caption{Values of cumulants for portfolios, the larger $j$ the portfolio 
	is supposed to be less ``variable''.}
\label{f::cu}
\end{figure}

In Fig.~(\ref{f::cu}) some cumulant value of the one dimensional random 
variable, that is 
the $j$'th investment portfolio $Y_j$ (with elements $y_{t,i} = \sum_{i=1}^{M} 
x_{t,i} V_{i,j}$) are 
presented for different methods of the factor matrix determination.
Generally large cumulants values were stored in first
portfolios where $j = 1, 2, \cdots$. For further investigation I took the $5$ 
rear portfolios, where $j \in [5,10]$ as those that have low cumulants' 
absolute values. 

\subsubsection{Testing optimal portfolios.}
After the training (the determination of $V$) has been completed, the testing 
of 
portfolios is performed. The factor matrices ($V$) columns contain both 
positive 
and negative values, the later corresponds to the negative value 
of shares in the portfolio -- the short sale. To diminish the use of the sort 
sale, 
the test portfolios were compared with the benchmark portfolio. Shares values
contributions in benchmark portfolio -- $BP_i$ are given in Tab.~(1).
In proposed test portfolios the value contribution of the $i$'th 
share in the $j$'th portfolio would be:  
\begin{equation}{\label{eq::test}}
TV_{i,j} = \frac{\alpha BP_i + V_{i,j}}{\sum_{i = 
			1}^{10} \left(\alpha BP_i + V_{i,j}\right)},
\end{equation}
the $\alpha 
= 7$ was taken, to make cases of the short sale rare.
For testing, shares prices of companies, see Tab.~(1) were 
taken. Testing set is represented by: 
${p_{t',i}}$, where $t'$ is time in the testing window. The percentage return 
of $j$'th portfolio after $L$ trading days is:
\begin{equation}
Pr_j(L) = 
\frac{\sum_{i=1}^{10} TV_{i,j}\left(\frac{p_{(t'=L+1,i)}}{p_{(t'=1,i)}}\right)  
- \sum_{i= 1}^{10}TV_{i,j}}{\sum_{i= 
1}^{10} TV_{i,j}} = \sum_{i=1}^{10} 
TV_{i,j}\left(\frac{p_{(t'=L+1,i)}}{p_{(t'=1,i)}} \right).
\end{equation}
In Fig.~(\ref{f::por}), returns after $10$ and $20$ trading days are 
presented. Remark, in this research transaction costs were 
not taken into account. The benchmark portfolio contributions can be reproduced 
by simply substituting $\forall_{i,j} \ V_{i,j} = 0$ to Eq.~(\ref{eq::test}).

\subsection{Discussion.} 
\begin{figure}
		\subfloat[After $10$ trading days]{
			\includegraphics{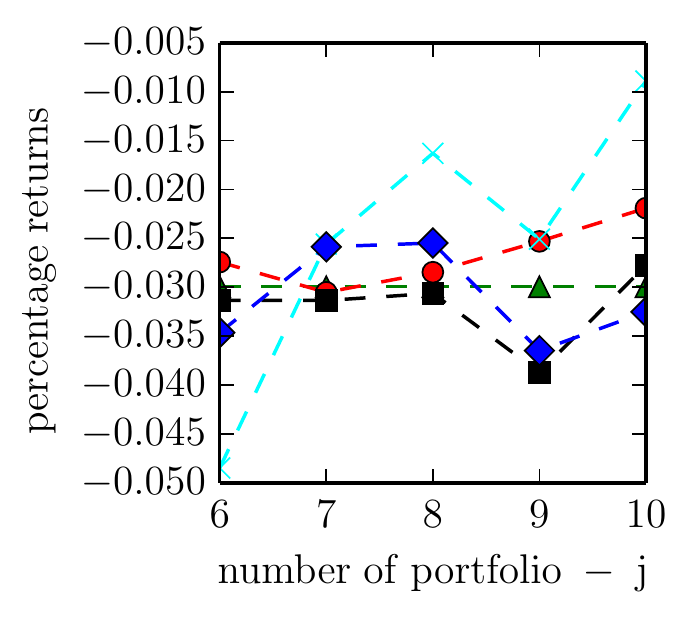}}
	\subfloat[After $20$ trading days]{
		\includegraphics{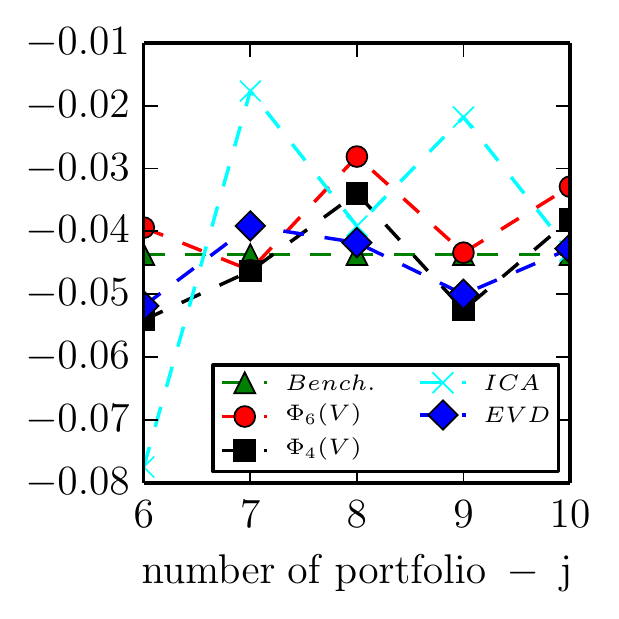}}
	\caption{Returns of $5$ portfolios, investments window 
	22.06.2015 -- 20.07.2015.}
	\label{f::por}
\end{figure}

Analysing Fig.~(\ref{f::por}), one can see that 
the $\Phi_6(V)$ method gives in both cases 
$3$ portfolios that are better than the benchmark and $2$ that are as good as 
benchmark. In the reminding part of the paper, I discuss the statistics of 
returns of such $5$ portfolios. 

\begin{figure}
	\subfloat[minimum of portfolio's returns]{\label{eq::min1}
	\includegraphics{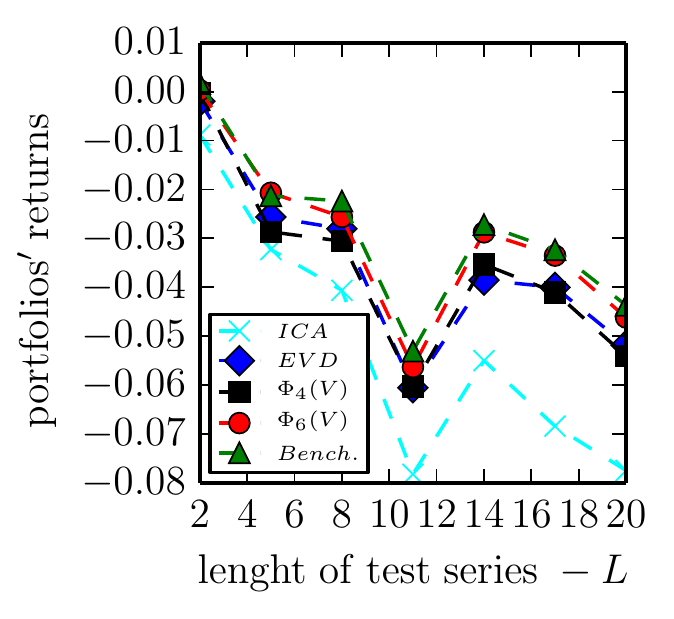}}
\subfloat[maximum of portfolio's returns]{\label{eq::max1}
	\includegraphics{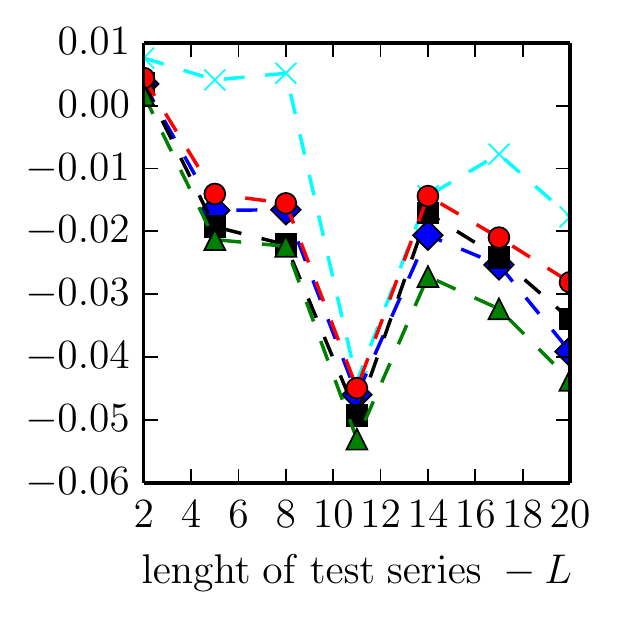}}\\
\subfloat[mean of portfolio's returns]{\label{eq::m1}
	\includegraphics{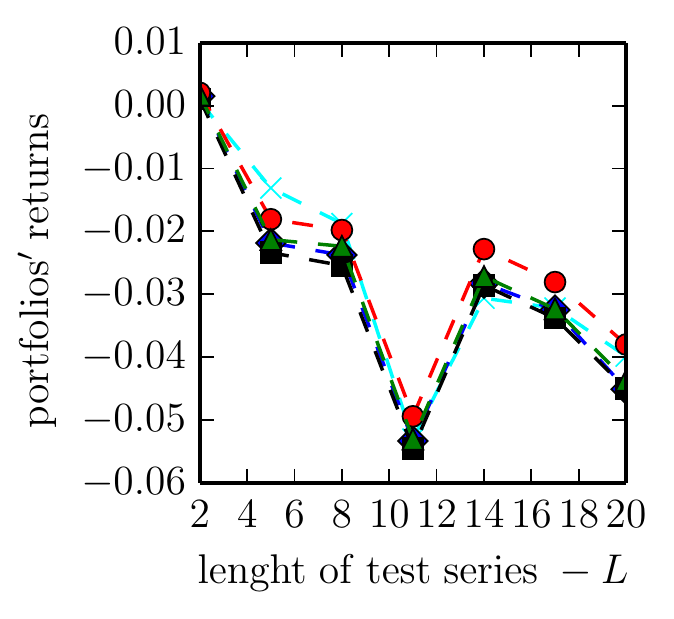}}
\subfloat[mode of portfolio's returns]{\label{eq::mo1}
		\includegraphics{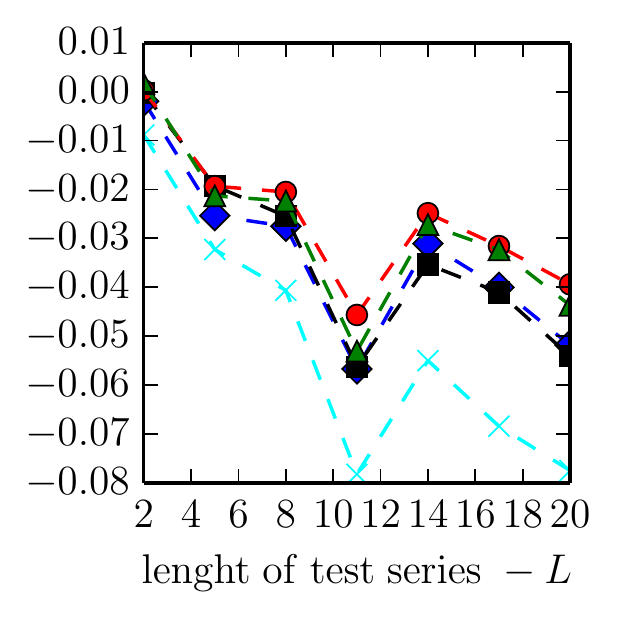}}
	\caption{Statistics of investment returns, $7$'th window of 
	22.06.2015 -- 20.07.2015.}
\label{eq::stat3}
\end{figure}

One can also conclude, that each method of factor matrix determination 
($\Phi_6(V)$, $\Phi_4(V)$, EVD, ICA) produces the worst 
portfolio
\begin{equation}
\min_{j \in [6,10]}Pr_j(L),
\end{equation}
which return is minimal and 
often smaller than benchmark's return. Those minimum of portfolios' returns are 
presented in Fig.~(\ref{eq::min1}). Analysing minimum of 
portfolios' returns one can 
conclude that out of all methods ($\Phi_6(V)$, $\Phi_4(V)$, EVD, ICA) the 
$\Phi_6(V)$ method gives smallest loss -- its worst portfolio is almost as good 
as benchmark. The worst results gives the ICA method, this is due to large 
variability of returns -- see Fig.~(\ref{f::por}), such method is not desirable 
during a crisis.
Similarly best portfolios can found:
\begin{equation}
\max_{j \in [6,10]}Pr_j(L),
\end{equation}
Their results are presented in Fig.~(\ref{eq::max1}). Best 
results presents the ICA, however this method is not safe. Next best are 
$\Phi_6(V)$ and $\Phi_4(V)$.

It is worth checking now, which portfolio determination method is best on 
average. In Fig.~~(\ref{eq::m1}), mean values of portfolios' 
returns are 
presented:
\begin{equation}
\frac{1}{5}\sum_{j=6}^{10}Pr_j(L).
\end{equation} 
 Remark that all methods but $\Phi_6(V)$ give an average return similar to or 
 worse than the benchmark. It is a worthy result, since it is hard to beat 
 the benchmark on average.
To examine a typical portfolio, the mode of portfolios' returns can  
be mentioned as well -- see Fig.~(\ref{eq::mo1}), here 
$\Phi_6(V)$ gives results, better than other methods, and slightly better the 
the benchmark. Concluding, statistics of the $\phi_6(V)$ method are better than 
other methods and the benchmark. Results of other $20$ days 
observation windows within the observation period determined by the Hurst 
exponent and outside it are discussed in next subsection.

\subsection{All observation windows.}
\begin{figure}
	\subfloat[whole investment 22.12.2014 -- 14.09.2015]{\label{eq::1w}
		\includegraphics{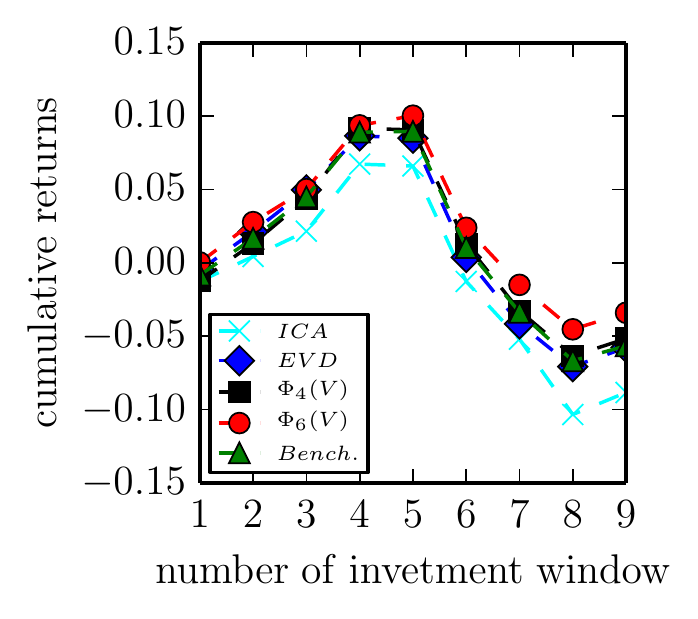}}
	\subfloat[crisis windows 23.04.2015 -- 14.09.2015]{\label{eq::1w1}
		\includegraphics{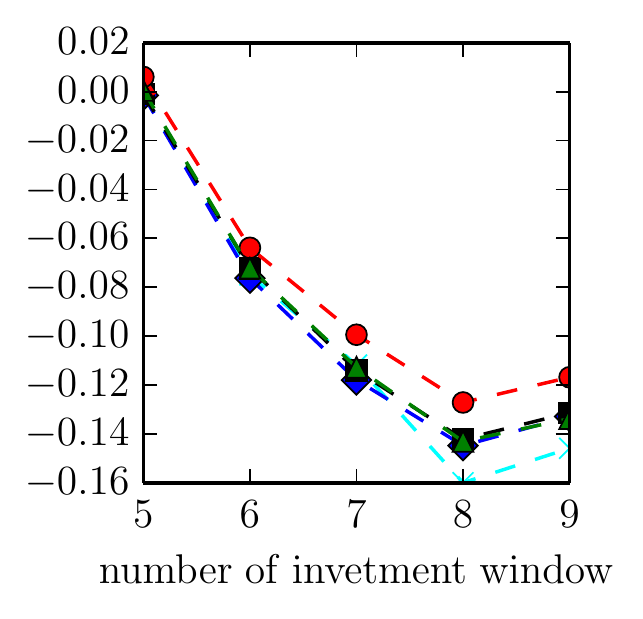}}
	\caption{Cumulative of means of portfolios returns.}
	\label{eq::stat5}
\end{figure}

\begin{figure}
	\subfloat[whole investment 22.12.2014 -- 14.09.2015]{\label{eq::1mo}
		\includegraphics{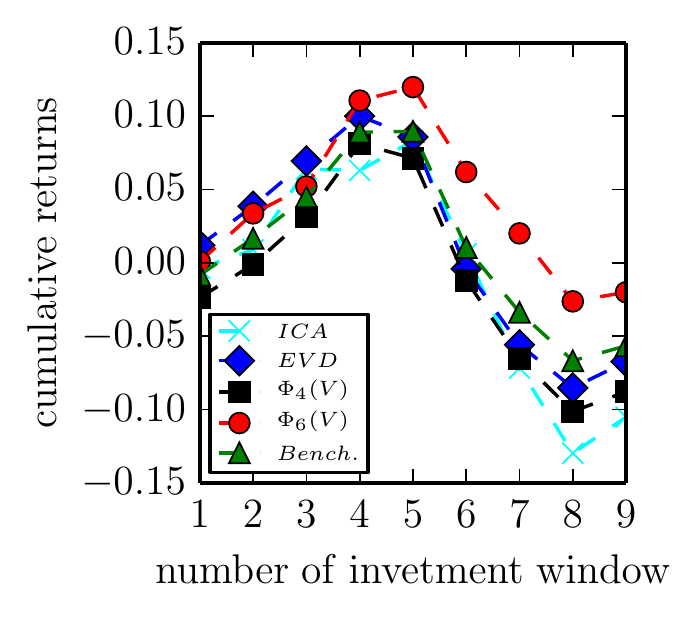}}
	\subfloat[crisis windows 23.04.2015 -- 14.09.2015]{\label{eq::1mo1}
		\includegraphics{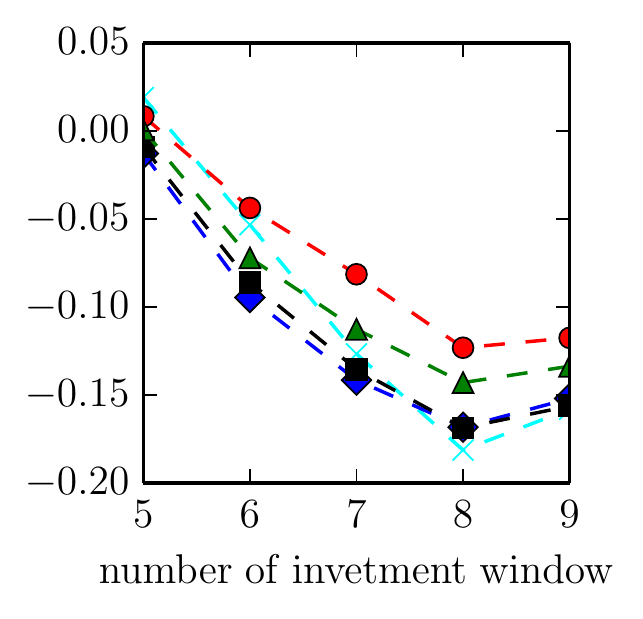}}
	\caption{Cumulative of modes of portfolios returns.}
	\label{eq::stat6}
\end{figure}

The analysis was performed for following observation windows: 22.12.2014 -- 
27.01.2015, 27.01.2015 -- 
24.02.2015, 24.02.2015 -- 24.03.2015, 24.03.2015 -- 23.04.2015, 23.04.2015 -- 
22.05.2015, 22.05.2015 -- 22.06.2015, 22.06.2015 -- 20.07.2015, 20.07.2015 -- 
17.08.2015, 17.08.2015 -- 14.09.2015. The first window starts a trading day 
after the enter signal recorded at 19.12.2014. The 
WIG20 index increased in first for windows, the maximum appeared in the $5$'th 
window where the crisis started, the last ($9$'th window) ends just after the 
exit signal recorded at 10.09.2015. Windows $5$ -- $9$ are crisis windows. 
Since I an interested in the investment strategy outcome between the enter and 
the exit signal, I present 
the cumulative results of investment that starts a trading day after the enter 
signal and ends just after exit signal. For each window factor matrices are 
calculated separately, investment is made at the first point in a window, at 
the last point of the window shares are sold and the mean of returns of $5$ 
portfolios is calculated. Cumulative of such mean returns are presented in 
Fig.~(\ref{eq::1w}). In Fig.~(\ref{eq::1w1}) the cumulative results are 
presented for crisis portfolios $5$ - $9$, here investment starts at 
23.04.2015.
In Fig.~(\ref{eq::stat6}) similar results are presented, but now mode of 
returns of $5$ portfolios is calculated in each window and the cumulative 
results are presented. Analysing 
Fig.~(\ref{eq::stat5}, \ref{eq::stat6}) one can conclude that the $\Phi_6(V)$
method on average gives best results at the exit point and during the crisis.

\begin{figure}
	\subfloat[Cumulatives of minima of returns]{\label{eq::min}
		\includegraphics{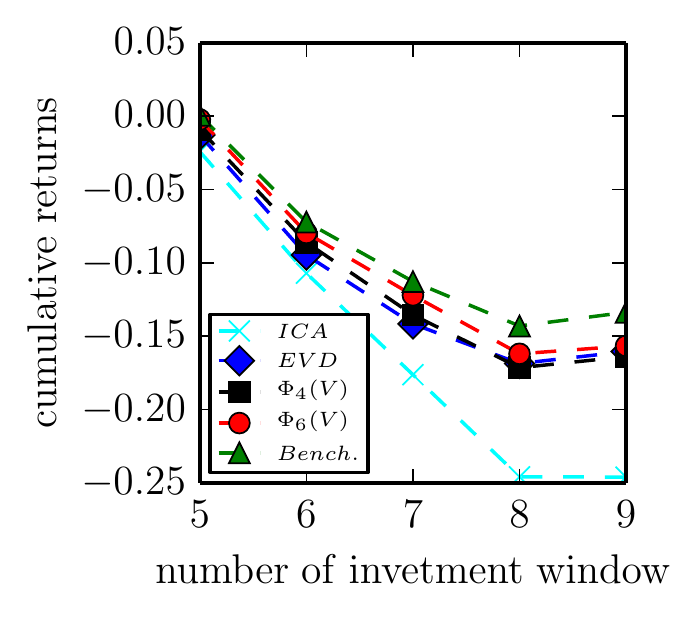}}
	\subfloat[Cumulatives of maxima of returns]{\label{eq::max}
		\includegraphics{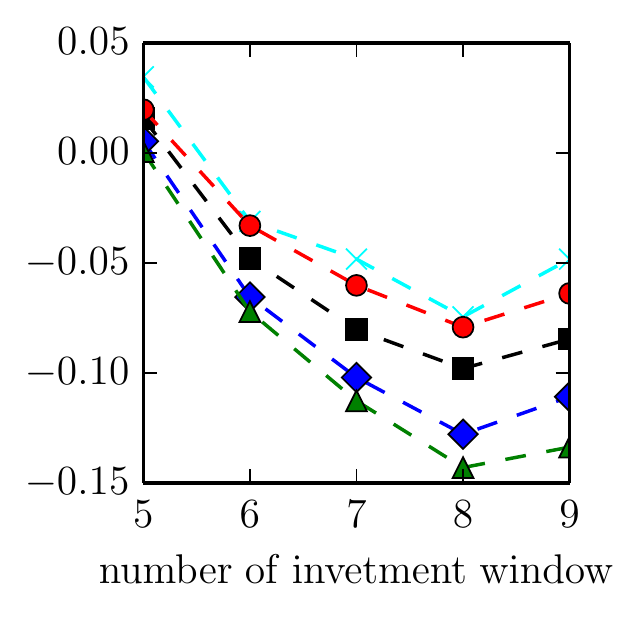}}
	\caption{Best and worst cumulative result at crisis windows 23.04.2015 -- 
	14.09.2015.}
	\label{eq::stat7}
\end{figure}

In Fig.~(\ref{eq::min}) cumulative results are presented, if in each window the 
worst portfolio was chosen (unlucky choice) -- there $\Phi_6(V)$ method is 
worse than a benchmark, but slightly better than other methods. In 
Fig.~(\ref{eq::max}) cumulative results are presented, if in each window the 
best portfolio was chosen (lucky choice) -- there $\Phi_6(V)$ method is better 
than 
all other methods apart from ICA. However the ICA produces also very bad 
portfolios (worst minimum), and hence is not adequate for a crisis.

\begin{figure}
	\subfloat[Cumulatives of means of returns]{\label{eq::mean}
		\includegraphics{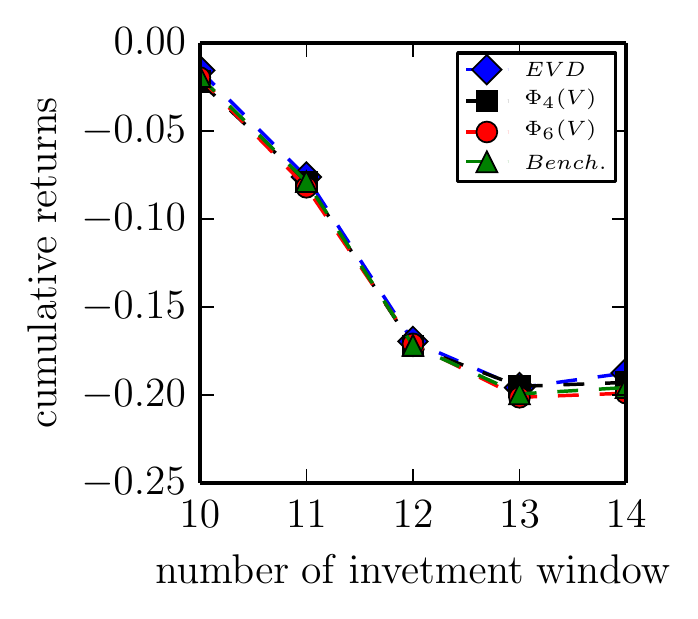}}
	\subfloat[Cumulatives of modes of returns]{\label{eq::mode}
		\includegraphics{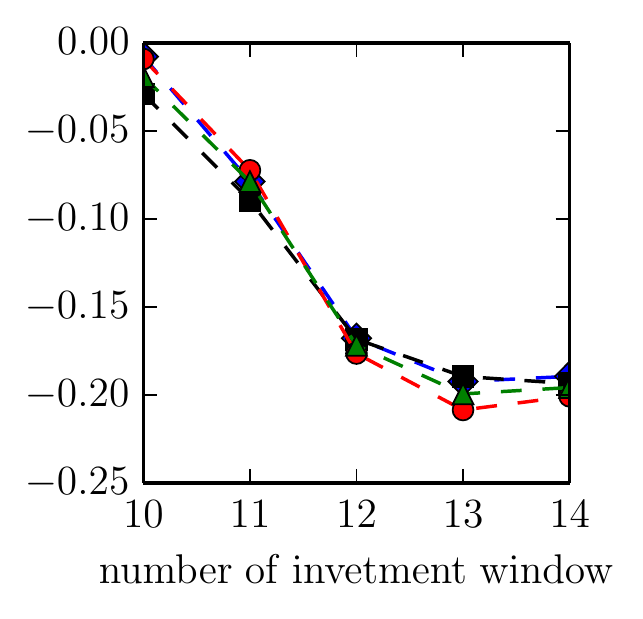}}
	\caption{Statistics for investment after exit signal, 14.09.2015 -- 
	09.02.2016}
	\label{eq::stat8}
\end{figure}

To test a method a bit more, I introduced observation windows after the exit 
signal, and number them as $10$'th to $15$'th, the cumulative results of means 
and modes of portfolio returns are presented in 
(\ref{eq::stat8}). Investment starts at 14.09.2015 and investment windows are 
14.09.2015 -- 12.10.2015, 12.10.2015 -- 09.11.2015, 
09.11.2015 -- 08.12.2015, 08.12.2015 -- 12.01.2016 and 12.01.2016 -- 09.02.2016.
It can be concluded that beyond the exit signal the $\Phi_6(V)$ method gives 
results similar to other methods and the benchmark. Hence the use of the Hurst 
exponent to determine the proper enter and exit signal appears to be crucial.

\section{Conclusions}
The author has used the multi--cumulant tensor analysis to 
analyse financial data and determine optimal investment portfolio with low 
absolute values of cumulants of their percentage returns. For this purpose, the 
author has analysed daily 
returns of shares traded on the Warsaw Stock 
Exchange to determine the factor matrix that represents such 
portfolios and test them during the recent rupture and crash period on the 
Warsaw Stock Exchange.

The main result of this 
work is the introduction of the algorithm that uses $2$'nd -- $6$'th cumulant 
tensors to analyse multivariate financial data and determine the investment 
portfolios that have low variability (low 
cumulants' absolute values). The Hurst exponent, calculated by the local DFA 
for the WIG20 index, indicates the auto--correlation phase on the stock 
market (the rupture period and the early stage of the crisis). At this phase, 
the introduced method is on average better than the benchmark and other tested 
methods. Importantly the Hurst exponent condition appears to be necessary to 
achieve this result. The examination of the 
method can be extended in further research, e.g. the algorithm can be tested on 
many stock exchanges. The algorithm can also be used to analyse other 
(non--financial) data that are 
non--Gaussian distributed.

\section*{Acknowledgements}
The research was partially financed by the National Science Centre, Poland - 
project number 2014/15/B/ST6/05204

\bibliographystyle{ieeetr}
\bibliography{finances}

\end{document}